\documentclass[a4paper,11pt]{article}
\pdfoutput=1 

\usepackage{jcappub} 

\usepackage[T1]{fontenc} 

\usepackage{newtxtext,newtxmath}

\title{Large scale structure constraints and matter power spectrum in $f (Q,\mathcal{L}_{m})$ gravity}

\author[a]{Praveen Kumar Dhankar,}
\author[b, c,1]{Albert Munyeshyaka,\note{Corresponding author.}}
\author[d]{Saddam Hussain,}
\author[e]{Tom Mutabazi}

\affiliation[a]{Symbiosis Institute of Technology, Nagpur Campus, Symbiosis International (Deemed University), Pune-440008, Maharashtra, India}
\affiliation[b]{Rwanda Astrophysics Space and Climate Science Research Group, University of Rwanda, College of Science and Technology, Kigali, Rwanda}
\affiliation[c]{Kibogora Polytechnic, Faculty of education, Western province, Rwanda}
\affiliation[d]{Institute of Theoretical Physics and Cosmology, Zhejiang University of Technology, Hangzhou 310023, China}

\emailAdd{pkumar6743@gmail.com}
\emailAdd{munalph@gmail.com}
\emailAdd{saddamh@zjut.edu.cn}

\abstract{In the present work, we take into account the dynamical system analysis to investigate the matter power spectrum within the framework of the $f(Q,\mathcal{L}_{m})$ gravitational theory. After obtaining autonomous dynamical system variables for two different particular pedagogical choices of $f(Q,\mathcal{L}_{m})$ models (A and B), we derive the full system of perturbation equations using the $1+3$ covariant formalism to study the matter fluctuations. We present and solve the energy density perturbation equations to obtain the energy density contrast, which decays with redshift for both models for a particular choice of model parameters. After obtaining the numerical results of the density contrast, we computed the matter spectra for each model and performed a comparative analysis with the $\Lambda$CDM. Furthermore, by employing Markov Chain Monte Carlo (MCMC) analysis, the model parameters were constrained using a combination of different observational data sets to improve the robustness and accuracy of the parameter estimation. Our results indicate that only model A can be compatible with the considered observational data sets.\\\\
\textit{keywords:} $f(Q, L_m)$ gravity; Hubble tension; MCMC analysis, Observation data.}

\begin{document}
\maketitle
\flushbottom

\section{Introduction}
The recent observations present convincing evidence that the current universe is experiencing an accelerated phase of expansion. Different studies have been trying to explain possible causes of this puzzle, though most of them face many challenges \cite{riess1998observational}.  The most acceptable model in trying to explain the cause of this acceleration is the concordance model dubbed the $\Lambda$CDM model, which utilizes the $\Lambda$ as a source of dark energy. This model agrees with a number of observations such as Type $I_{a}$ supernovae \cite{riess1998observational}, cosmic microwave background (CMB) \cite{giostri2012cosmic,komatsu2009five,jarosik2011seven}, Baryon acoustic oscillations (BAO) \cite{weinberg2013observational,delubac2013baryon} and large scale structure \cite{daniel2008large}. Despite the remarkable successes, this model faces different challenges in the early and late time universe such as the fine tuning problems, the problems of Hubble and $\sigma 8$ tensions. This has led to a search for alternative theory to explain the possible source of dark energy and in line with observational predictions. Among the leading theory includes the consideration of the geometrical source of dark energy which leads to the consideration of modified theories of gravity such as $f(R)$ \cite{capozziello2007newtonian,ilyas2024stability}, $f(G)$ \cite{sharif2019ghost,lohakare2024cosmology}, $f(T)$ \cite{myrzakulov2011accelerating} and  $f(Q)$ \cite{heisenberg2024review,mandal2020energy} to name but a few, where $R$ is the Ricci scalar, $G$  is the Gauss-Bonnet invariant, $T$ is the torsion scalar and $Q$ is the nonmetricity scalar, respectively.\\ The simplest theory $f(R)$ arises through an extension of the Einstein-Hilbert action by replacing Ricci scalar by an arbitrary function for the Ricci scalar to form the $f(R)$ theory. This theory is able to explain the cosmic accelerated expansion in geometrical way without the need of dark energy and it can explain the flat rotational curves of galaxies \cite{song2007large,bohmer2008dark,berry2011linearized,chiba2007solar,guo2014solar}. However the $f(R)$ theory of gravity faces some challenges for example when subjected to solar system test, and matter instability issues \cite{sokoliuk2023impact,motohashi2015third}. \\An alternative method to extending the Einstein-Hilbert  action involves postulating the presence of a non-minimal coupling, for example between geometry and matter leading to $f(R,\mathcal{L}_{m})$ gravity, geometry and the trace of energy-momentum tensor giving rise to $f(R,T)$ gravity \cite{harko2010f,wang2012energy,jaybhaye2023baryogenesis,maurya2023constrained,singh2023constrained,myrzakulov2012frw,bhattacharjee2020inflation,shabani2017stability}. In this case gravitational dynamics is described by more general functions of the curvature scalar, matter Lagrangian and the trace of energy momentum tensor, respectively allowing to investigate gravitational behaviors beyond GR and $\Lambda$CDM predictions. \\The $f(T)$ gravity theories have been extensively explored in different contexts and distinguished among theories which can explain the cosmic acceleration without the need for dark energy \cite{mishra2025exploring,bahamonde2020solar,capozziello2011cosmography,li2011large}. Another alternative theory in a geometrical formulation of gravitation includes the use of nonmetricity $Q$ of the metric explored in different contexts, for examples the work conducted in \cite{sokoliuk2023impact} explored the impact of $f(Q)$ gravity on the large scale structure and the authors investigated the logarithmic $f(Q)$ gravity in terms of observational constraints using Monte-Carlo-Markov-Chain (MCMC) method and high body simulations, whereas in \cite{wang2025solar}, the authors discussed the Solar system tests in covariant $f (Q)$ gravity. This theory offers an advantage of coordinate calculations in GR in a covariant way. Cosmology of $f(Q)$ gravity theory and its observational constraints were analysed in \cite{mandal2023cosmological,koussour2023observational,ayuso2021observational,lazkoz2019observational}.  A key aspect of $f(Q)$ gravity theory is that it is possible to distinguish gravity from inertial effects. This theory offers a straightforward formulation in which self-accelerating solutions spontaneously appear in both early and late time universes and has a substantial benefit that the background field equations are always second order and the matter instabilities can be avoided in the $f(Q)$ theory \cite{motohashi2015third,sokoliuk2023impact}.\\ The $f(Q)$ gravity theory has been explored in a number of studies which confirmed its authenticity in reproducing both the early and the late time dynamics of the universe in a way similar to the $\Lambda$CDM model. It can satisfy constraints from different observational probes  such as CMB, Supernovae type $I_{a}$ distance modulus, BAO and  Observational Hubble Data set (OHD).  \\ Nevertheless, there exists  different types of observations that are  sensitive not only to the expansion history of the universe but also to the evolution of matter perturbations. The fact that the evolution of perturbations depends on the specific gravity theory has made these observations a key actor in distinguishing  between different cosmological models and the mechanism driving cosmic acceleration \cite{sahlu2025structure}. In this regard, as far as cosmological perturbation is concerned, the $1+3$ covariant perturbations explored extensively in previous works \cite{sahlu2025structure,sami2021covariant} can be used to obtain energy density perturbation equations. Different works have been considering this formalism in the context of $f(Q)$ gravity to investigate the evolution of large-scale structure. For example, the work conducted in \cite{jimenez2020cosmology} investigated cosmological perturbations in $f(Q)$ gravity and discussed the re-scaling of Newton constant in tensor perturbations. The scalar parts of the perturbations introduced two additional propagating modes, suggesting that $f(Q)$ gravity presents at least two extra degrees of freedom. \\ Moreover, extending non-metric gravity by incorporating the trace of energy-momentum tensor $T$ of matter to formulate  $f(Q, T)$ has been investigated in different works. The $f(Q,T)$ theory was observationally constrained and some of the models of the $f(Q,T)$ gravity have successfully described  accelerated expansion of the universe \cite{xu2019f,xu2020weyl,arora2020f,arora2021constraining}, offering viable alternatives to the standard $\Lambda$CDM paradigm. \\ The work conducted by \cite{hazarika2024f} extended the non-metric gravity by incorporating the matter Lagrangian in the Lagrangian density of the $f(Q)$ gravity to obtain $f(Q,\mathcal{L}_{m})$ gravity theory. The authors among other things investigated the conservation problem of the matter energy momentum tensor, investigated the cosmic evolution for the case of a flat Friedmann-Lemaitre-Robertson-Walker metric and further consider two particular gravitational models of in the context of $f(Q,\mathcal{L}_{m})$ gravity. The authors compared the evolution of gravitational models with the predictions of two distinct observational data sets.\\
In the work conducted in \cite{myrzakulov2024late}, the authors investigated late-time cosmology in the context of $f(Q,\mathcal{L}_{m})$ gravity and discussed its analytical solutions and observational fits. The authors showed that this theory of gravity contributes to our understanding of the universe's expansion dynamics.\\ In \cite{myrzakulov2025constraining}, $f(Q,\mathcal{L}_{m})$ gravity was constrained with bulk viscosity, and the findings underscore the significant role of bulk viscosity in understanding accelerated expansion in the universe within alternative gravity theories.\\
The work carried out in \cite{myrzakulov2024observational} investigated the observational late-time acceleration in $f(Q,\mathcal{L}_{m})$ gravity, and their analysis showed that the $f(Q,\mathcal{L}_{m})$ model aligns well with the observational results and exhibits similar behaviours to the $\Lambda$CDM model. Motivated by these works, the  present work  aims to extend the work conducted in \cite{hazarika2024f} to perturbation level  to investigate the implications of the considered gravitational models on the evolution of large-scale structure and on the matter power spectrum. In this regards, we use dynamical system approach to obtain autonomous ordinary differential  (ODE). We then use the $1+3$ covariant formalism to obtain the energy density perturbation equations. Solving both the autonomous ODEs and the perturbation equations we compute the matter power spectra for each model. The results inform that the effects of the gravitational models considered in this work on the matter power spectra within this $f(Q,\mathcal{L}_{m})$ gravity theory provide promising results on the evolution of large scale structure. The next aim is to use MCMC analysis to constrain model parameters resulting from the use of $f(Q,\mathcal{L}_{m})$ gravity theory. In so doing, we compute the corner (triangular) plots and present the mean value parameters corresponding to the combinations of data sets namely: (i) CC+PP (ii) CC+DES. For further analysis, we use combinations of  redshift space distortion data $f\sigma_{8}$ data with the latest measurements of the growth rate and amplitude of matter fluctuations $\sigma_{8}$ namely:  (iii) CC+BAO+fs8 (iv) CC+BAO+fs8+PP (v) CC+BAO+fs8+DES. Using the MCMC simulations, we are able to constrain the best fit model parameters in the context of $f(Q,\mathcal{L}_{m})$ gravity theory.
\\The rest of this paper is organized as follows: In Section (\ref{sec2}), we present the mathematical framework, where cosmological equations are discussed in the context of $f(Q,\mathcal{L}_{m})$ gravity and the dynamical system equations for the defined models. In Section (\ref{sec3}), we present the covariant density perturbation equations and their corresponding numerical results (energy density contrast). Section (\ref{sec4})  computes and presents the matter power spectra resulting from the energy density perturbation equations and the dynamical variables of the system for the $f(Q,\mathcal{L}_{m})$ gravity models considered, while Section (\ref{sec5}) presents the structure growth equation. In Section (\ref{sec:resutls}), we present the data used to constrain model parameters and the results obtained, whereas Section (\ref{sec7}) discusses the results and concludes the work.
\section{Background and perturbation equations in $f(Q,\mathcal{L}_{m})$ gravity}\label{sec2}
This section provides an overview of the geometric foundations of the $f(Q,\mathcal{L}_{m})$ gravitational theory. We present an action representing this theory, the corresponding gravitational field equations and the conservation equations of the energy momentum tensor. The action principle helps to control the dynamics of a physical system. In the $f(Q,\mathcal{L}_{m})$ context, the action is represented as \cite{myrzakulov2024late,myrzakulov2025constraining,hazarika2024f}
\begin{equation}
	S=\int \sqrt{-g}f\Big(Q,\mathcal{L}_{m}\Big)d^{4}x,\label{eq2.1}
\end{equation} where $\sqrt{-g}$ is the determinant of the metric $g_{\mu\nu}$ and $f(Q,\mathcal{L}_{m})$ is an arbitrary function of nonmetricity scalar $Q$ and  of matter Lagrangian $\mathcal{L}_{m}$. The nonmetricity scalar $Q$ is given by $Q=-Q_{\lambda \mu \nu}P^{\lambda \mu \nu}$ and it describes the deviation of the manifold geometry from isotropy and can be thought of as a measure of how much the volume of a parallelly transported object changes as it moves through space-time, where $P^{\lambda \mu \nu}$ is a super-potential--a nonmetricity conjugate \cite{myrzakulov2024late,hazarika2024f}. By varying the action (eq. \ref{eq2.1}) with respect to the metric tensor and $Q$ and by applying the boundary conditions, integrating and equating the metric variation of the action to zero, one can get the field equations of $f(Q,\mathcal{L}_{m})$ gravity as 
\begin{eqnarray}
	&&\frac{2}{\sqrt{-g}}\bigtriangledown_{\alpha}\Big(f_{Q}\sqrt{-g}P^{\alpha}_{\mu \nu}\Big)+f_{Q}\Big(P_{\mu \alpha \beta}Q^{\alpha \beta}_{\nu}-2Q^{\alpha \beta}_{\mu}P_{\alpha \beta \nu}\Big)\nonumber\\&&+\frac{1}{2}f g_{\mu \nu}=\frac{1}{2}f_{lm}\Big(g_{\mu \nu}\mathcal{L}_{m}-T_{\mu \nu}\Big),
\end{eqnarray} where $T_{\mu \nu}=-\frac{2}{\sqrt{-g}}\frac{\delta (\sqrt{-g}\mathcal{L}_{m})}{\delta g^{\mu \nu}}$ is the energy-momentum tensor of the matter, $\delta \sqrt{-g}=-\frac{1}{2}\sqrt{-g}g_{\mu \nu}\delta g^{\mu \nu}$ is the variation of the determinant of the metric and $\delta Q=2P_{\alpha \nu \rho}\bigtriangledown^{\alpha}\delta g^{\nu \rho}-\Big(P_{\mu \alpha \beta}Q^{\alpha \beta}_{\nu}-2Q^{\alpha \beta}_{\mu}P_{\alpha \beta \nu}\Big)\delta g^{\mu \nu}$ is the variation of $Q$. $f_{Q}=\frac{\partial f(Q,\mathcal{L}_{m})}{\partial Q}$ and $f_{lm}=\frac{\partial f(Q,\mathcal{L}_{m})}{\partial \mathcal{L}_{m}}$. For the case $f(Q,\mathcal{L}_{m})=f(Q)+2\mathcal{L}_{m}$, it reduces to the field equations of $f(Q)$ gravity \cite{jimenez2018coincident}.  Consider  matter as a perfect fluid, the energy-momentum tensor is presented as 
\begin{equation}
	T^{\mu}_{\nu}=\Big(\rho+p\Big)u_{\nu}u^{\mu}+pg^{\mu}_{\nu},
\end{equation} where $\rho$ is the energy density and $p$ is the pressure. $u^{\mu}$ denotes the four-velocity of the fluid. In this case, the conservation equation can be written by 
\begin{equation}
	\dot{\rho}+3H\Big(\rho+p\Big)=0.
\end{equation}
In order to explore the cosmological evolution of the Friedmann-Robert-Walker (FRW) universe in $f(Q,\mathcal{L}_{m})$ gravity, let us   consider a flat geometry and the space-time metric given by
\begin{equation}
	dS^{2}=-dt^{2}+a(t)^{2}\Big(dx^{2}+dy^{2}+dz^{2}\Big),
\end{equation} where $a(t)$ is the scale factor governing the expansion of the universe, which is related to the Hubble parameter $H$ as $H=\frac{\dot{a}}{a}$. The dot denotes derivative with respect to cosmic time. The nonmetricity is given by $Q=6H^{2}$.  The non-zero components of the energy-momentum tensor are therefore given by $T^{\mu}_{\nu}=\Big(\rho,p,p,p\Big)$. In the following subsection, we present the Friedmann equations in the context of $f(Q,\mathcal{L}_{m})$ gravitational theory. 
\subsection{Generalised Friedmann equations}
Considering the Friedmann-Robert-Walker (FRW) metric in a flat geometry, the Friedmann equations and the continuity equation can be presented , respectively as
\begin{eqnarray}
	&&3H^{2}=\rho_{eff},\label{eq2.6}\\
	&&2\dot{H}+3H^{2}=-p_{eff},\label{eq2.7}\\
	&&\dot{\rho}_{eff}+3H\Big(\rho_{eff}+p_{eff}\Big)=0\label{eq2.8},
\end{eqnarray} where 
\begin{eqnarray}
	&&\rho_{eff}=\frac{1}{4f'}\Big[f-f_{lm}\Big(\rho+l_{m}\Big)\Big],\label{eq2.9}\\
	&&p_{eff}=2\frac{\dot{f}'H}{f'}-\frac{1}{4f'}\Big[f+f_{l_{m}}\Big(\rho+2p-l_{m}\Big)\Big].\label{eq2.10}
\end{eqnarray} 

By looking at eq. (\ref{eq2.6}) and eq. (\ref{eq2.7}), the cosmological evolution equations contain extra-terms representing the contributions from non-nonmetricity-matter coupling leading to geometrical dark energy. These terms drive the recent accelerated expansion of the universe.
\subsection{Cosmological models}
In this part, we investigate two different cosmological models namely model A and model B given by $f=-\alpha Q+2l_{m}+\beta$ and $f=-\alpha Q+(2l_{m})^{2}+\beta$, respectively explored in  \cite{myrzakulov2025constraining,hazarika2024f}. In these works the authors assumed that matter in the universe obeys an equation of state given by $p=(\gamma-1)\rho$, with $1\leq \gamma \leq2$. For $\gamma=\frac{4}{3}$, the equation of state describes the radiation dominated universe, early universe (high density), whereas for $\gamma=1$, it describes the pressure-less matter (dust). 
\subsubsection{Model A}
Consider the model A given by $f=-\alpha Q+2l_{m}+\beta$, with $l_{m}=p$ and $p=\Big(\gamma-1\Big)\rho$, the Friedmann equations (eq. \ref{eq2.6} to eq. \ref{eq2.10}) reduces to
\begin{eqnarray}
	&& 3H^{2}=-\frac{\beta}{2\alpha}+\frac{\rho}{\alpha},\label{eq2.11}\\
	&&2\dot{H}+3H^{2}=-3H^{2}\Big(\gamma-1\Big)-\frac{\beta \gamma}{2\alpha},\label{eq2.12}\\
	&&\rho_{eff}=\frac{\rho}{2\alpha}-\frac{\beta}{2\alpha},\\
	&&p_{eff}=3H^{2}\Big(\gamma-1\Big)+\frac{\beta \gamma}{2\alpha},\\
	&&H(z)=\Big[\frac{\Big(6H^{2}_{0}\alpha+\beta\Big)\Big(1+z\Big)^{3\gamma}-\beta}{6\alpha}\Big]^{\frac{1}{2}}.\label{eq2.15}
\end{eqnarray}

In the work conducted by \cite{hazarika2024f}, the authors focused mainly on the variables as a function of the redshift of the energy density, deceleration parameter and the effective of equation of state and used MCMC simulations to constrain model parameters. However, in the present we will use the same models to explore the dynamical system and the $1+3$ covariant perturbations approaches to investigate the effect of each model on large scale structure formation and matter power spectrum. Starting from  Eq. (\ref{eq2.11}) and eq. (\ref{eq2.12}), we  can  find the dynamical system variables.
\subsubsection{Dynamical system variables for model A}
In the present section, we present the dynamical system variables \cite{khyllep2021cosmological,munyeshyaka1,ntahompagaze2022large,abebe2013large} in the context of $f(Q, \mathcal{L}_{m})$ gravity for model A.
From eq. (\ref{eq2.11}) we can rewrite
\begin{eqnarray}
	&&1=-\frac{\beta}{6\alpha H^{2}}+\frac{\rho}{3\alpha H^{2}},\label{eq2.16}\\
	&& 1+x-\Omega_{1}-\Omega_{2}=0,\label{eq2.17}\\
	&&x=\frac{\beta}{6\alpha H^{2}},\label{eq2.18}\\
	&&\Omega_{1}=\frac{\rho_{m}}{3\alpha H^{2}},\label{eq2.19}\\
	&&\Omega_{2}=\frac{\rho_{r}}{3\alpha H^{2}}\label{eq2.20}.
\end{eqnarray}
Where $\rho=\rho_{m}+\rho_{r}$, $\Omega_{1}=\frac{\Omega_{m}}{\alpha}$ and $\Omega_{2}=\frac{\Omega_{r}}{\alpha}$. From eq. (\ref{eq2.12}), we can get
\begin{equation}
	\frac{\dot{H}}{H^{2}}=-\frac{3}{2}\gamma \Big(x+1\Big).\label{eq2.21}
\end{equation} Using eq. (\ref{eq2.21}) and the fact that $\frac{df}{Hdt}=\frac{df}{dN}$, equations (\ref{eq2.18}), (\ref{eq2.19}) and (\ref{eq2.20}) evolve as
\begin{eqnarray}
	&&\frac{dx}{dN}=-3\gamma x-3\gamma x^{2},\label{eq2.22}\\
	&&\frac{d \Omega_{1}}{dN}=3\Omega_{1}\Big(\gamma-1\Big)+3\gamma \Omega_{1} x\label{eq2.23},\\
	&&\frac{d \Omega_{2}}{dN}=\Omega_{2}\Big(3\gamma-4\Big)+3\gamma \Omega_{2} x,\label{eq2.24}\\
	&&\frac{dh}{dN}=3h\Big(\Omega_{2}+\frac{\Omega_{1}}{2}\Big),\label{eq2.25}
\end{eqnarray} 
where $h=\frac{H}{H_{0}}$, $\dot{\rho}_{m}=-3H\rho_{m}$ and $\dot{\rho}_{r}=-4 H \rho_{r}$. The equations (\ref{eq2.22}),  (\ref{eq2.23}), (\ref{eq2.24}) and (\ref{eq2.25}) can be represented in redshift using $\frac{df}{dN}=-(1+z)\frac{df}{dz}$ as
\begin{eqnarray}
	&&-(1+z)\frac{dx}{dz}=-3\gamma x-3\gamma x^{2},\label{eq2.26}\\
	&&-(1+z)\frac{d \Omega_{1}}{dz}=3\Omega_{1}\Big(\gamma-1\Big)+3\gamma \Omega_{1} x\label{eq2.27},\\
	&&-(1+z)\frac{d \Omega_{2}}{dz}=\Omega_{2}\Big(3\gamma-4\Big)+3\gamma \Omega_{2} x,\label{eq2.28}\\
	&&-(1+z)\frac{dh}{dz}=3h\Big(\Omega_{2}+\frac{\Omega_{1}}{2}\Big).\label{eq2.29}
\end{eqnarray} 
Eqs. (\ref{eq2.26})--(\ref{eq2.29}) form an autonomous system of ordinary differential equations of the $f(Q, \mathcal{L}_{m})$ gravity in redshift space for the model A. The dimensionality of this system of equations can be reduced using the Friedmann constraint (eq. \ref{eq2.17}). The evolution of the Hubble parameter (H) was determined using eq. (\ref{eq2.11}) in terms of dynamical system variables. The stability analysis of dynamical variables in the context of $f(Q)$ gravity was conducted in \cite{khyllep2021cosmological}. In this work, the authors showed points corresponding to a deceleration-radiation dominated universe, a deceleration-matter dominated universe and an accelerated dark energy dominated universe. Ref. \cite{an2016dynamical} discussed the dynamical analysis of modified gravity with nonminimal gravitational coupling to matter, whereas Ref. \cite{paliathanasis2015dynamical} conducted the dynamical analysis in scalar field cosmology. This system of equations (eq. (\ref{eq2.26})--(\ref{eq2.29})) together with the energy density perturbation equations (which will be presented in the next sections) will be used to compute the matter power spectrum for model A in $f(Q, \mathcal{L}_{m})$ gravity. In the next subsection, let us derive dynamical system equations for the second model dubbed model B
\subsubsection{Model B}
Considering model B defined as $f=-\alpha Q+(2l_{m})^{2}+\beta$, the Friedmann equations (eq. \ref{eq2.6}--\ref{eq2.10}) and the Hubble parameter in the $f(Q, \mathcal{L}_{m})$ gravity can be modified as \cite{hazarika2024f}
\begin{eqnarray}
	&& 3H^{2}=-\frac{2}{\alpha}\Big(1-\gamma^{2}\Big)\rho^{2}-\frac{\beta}{2\alpha},\label{eq2.30}\\
	&&2\dot{H}+3H^{2}=\frac{\Big(\beta+6\alpha H^{2}\Big)\Big(\gamma-1\Big)}{2\alpha\Big(\gamma+1\Big)}-\frac{\beta }{2\alpha},\label{eq2.31}\\
	&&H(z)=\Big[\frac{\Big(6H^{2}_{0}\alpha+\beta\Big)\Big(1+z\Big)^{\frac{6\gamma}{1+\gamma}}-\beta}{6\alpha}\Big]^{\frac{1}{2}}\label{eq2.32}
\end{eqnarray}
The dynamical system variables for model B can be obtained by first rewriting  the above equation (from eq. \ref{eq2.30}) as
\begin{eqnarray}
	&&1=-\frac{\beta}{6\alpha H^{2}}+\frac{2\Big(\gamma^{2}-1\Big)\rho^{2}}{3\alpha H^{2}},\\
	&& 1+y-\Omega_{3}-\Omega_{4}=0,\label{eq2.34}\\
	&&y=\frac{\beta}{6\alpha H^{2}},\label{eq2.35}\\
	&&\Omega_{3}=\frac{2\Big(\gamma^{2}-1\Big)\rho_{m}^{2}}{3\alpha H^{2}},\label{eq2.36}\\
	&&\Omega_{4}=\frac{2\Big(\gamma^{2}-1\Big)\rho_{r}^{2}}{3\alpha H^{2}}\label{eq2.37}.
\end{eqnarray}
Where $\rho=\rho_{m}+\rho_{r}$, $\Omega_{3}=2\frac{\Big(\gamma^{2}-1\Big)}{\alpha}\Omega_{m}$ and $\Omega_{4}=2\frac{\Big(\gamma^{2}-1\Big)}{\alpha}\Omega_{r}$. 

From eq. (\ref{eq2.31}), the ratio of time derivative of $H$ and the square of $H$ can be calculated as
\begin{equation}
	\frac{\dot{H}}{H^{2}}=-\frac{3}{\gamma+1}-\frac{3y}{\gamma+1}.\label{eq2.33}
\end{equation}
Using eq. (\ref{eq2.30}) and the fact that $\frac{df}{Hdt}=\frac{df}{dN}$, the evolution of the dynamical system variables (eq. (\ref{eq2.35}), eq. (\ref{eq2.36}) and Eq. (\ref{eq2.37})) can be presented as  
\begin{eqnarray}
	&&\frac{dy}{dN}=\frac{6y}{\gamma+1} +\frac{6y^{2}}{\gamma+1},\label{eq2.39}\\
	&&\frac{d \Omega_{3}}{dN}=-3\Omega_{3}\Big(\frac{1+2\gamma}{\gamma+1}\Big)+\frac{3\Omega_{3} y}{\gamma+1},\label{eq2.40}\\
	&&\frac{d \Omega_{4}}{dN}=-\frac{2\Big(1+4\gamma\Big)}{\gamma+1}+\frac{3 \Omega_{4} y}{\gamma+1},\label{eq2.41}\\
	&&\frac{dh}{dN}=3h\Big(2\Omega_{4}+\Omega_{3}\Big)\label{eq2.42},
\end{eqnarray}
which can be rewritten in redshift space as 
\begin{eqnarray}
	&&-(1+z)\frac{dy}{dz}=\frac{6y}{\gamma+1} +\frac{y^{2}}{\gamma+1},\label{eq2.43}\\
	&&-(1+z)\frac{d \Omega_{3}}{dz}=-3\Omega_{3}\Big(\frac{1+2\gamma}{\gamma+1}\Big)+\frac{3\Omega_{3} y}{\gamma+1},\label{eq2.44}\\
	&&-(1+z)\frac{d \Omega_{4}}{dz}=-\frac{2\Big(1+4\gamma\Big)}{\gamma+1}+\frac{3 \Omega_{4} y}{\gamma+1},\label{eq2.45}\\
	&&-(1+z)\frac{dh}{dz}=3h\Big(2\Omega_{4}+\Omega_{3}\Big)\label{eq2.46}.
\end{eqnarray}
Eqs. (\ref{eq2.43})--(\ref{eq2.46}) form an autonomous system of ordinary differential equations in the redshift space for model B in $f(Q, \mathcal{L}_{m})$ gravity. These equations together with the energy density perturbations will be used to analyse the effect of this model on structure formation and on the matter power spectrum. In the next section, we derive the energy density perturbation equations in the context of $f(Q, \mathcal{L}_{m})$ gravity.
\section{Energy density perturbation in $f(Q, \mathcal{L}_{m})$ gravity}\label{sec3}
It is currently  a well known fact that the universe is not perfectly smooth, but full of large scale structure such as galaxies, clusters of galaxies and voids, to name but a few. These structures are believed to be the result of primordial fluctuations. Cosmological perturbations provide a room for explaining how structures grow from small fluctuations to large structures we see today in the universe.  There are currently two approaches to perturbation, namely, the metric-based approach, developed by Lifshitz \cite{lifshitz1946gravitational}, Badeen \cite{bardeen1980gauge}, and the covariant approach developed by Hawking \cite{hawking1966perturbations} and Ellis and Bruni \cite{ellis1989covariant}. These approaches differ in the fact that it is difficult in dealing with non-linear theory and requires that the metric be specified from the start for the metric approach, whereas the covariant formalism helps in describing space-time via covariantly defined variables with respect to a partial frame, such as the $1+3$ covariant space-time decomposition techniques.  It is a suitable method to describe physics and geometry using tensor quantities and relations valid in all coordinate systems. Nonlinearities can be accommodated, but the main advantage  of the $1+3$ covariant approach is that no unphysical gauge modes exist \cite{sahlu2020scalar,sami2021perturbations,abebe2023perturbations,munyeshyaka20231+,munyeshyaka2024covariant,munyeshyaka2023perturbations,ntahompagaze20251+}. Previous work in the context of $f(Q, \mathcal{L}_{m})$ gravity focused on the study of cosmological dynamics of the universe on the background expansion history \cite{hazarika2024f} and other works considered perturbation part in the context of $f(Q)$ gravity \cite{sahlu2025structure}. To our knowledge, there is no work in the literature on large scale structure formation scenario and matter power spectrum within the framework of $f(Q, \mathcal{L}_{m})$ gravity for  two different considered  models in this work. We therefore fill this gap by studying the energy density perturbation and matter power spectrum of these models using the $1+3$ covariant formalism. To do so, we first define the covariant and gauge-invariant gradient variables that describe perturbation in the matter density, expansion, the nonmetricity scalar and its energy momentum as \cite{munyeshyaka2023perturbations,sahlu2025structure} 
\begin{eqnarray}
	&& D^{m}_{a}=\frac{a\tilde{\bigtriangledown}_{a}\rho_{m}}{\rho_{m}},
	Z_{a}=a\tilde{\bigtriangledown}_{a}\theta,
	\mathcal{C}_{a}=a\tilde{\bigtriangledown}_{a}Q,
	\mathcal{F}_{a}=a\tilde{\bigtriangledown}_{a}\dot{Q}.\label{eq2.47}
\end{eqnarray}
The terms $D^{m}_{a}$ and $Z_{a}$ represent the energy density and the volume expansion of the fluid, respectively , and they are the basic tools to extract the evolution equations for matter fluctuations. Based on nonmetricity fluid, the terms $\mathcal{C}_{a}$ and $\mathcal{F}_{a}$ represent the spatial gradients of gauge-invariant quantities characterising the fluctuations in the nonmetricity density and momentum, respectively. By applying the technic commonly used in the $1+3$ covariant formalism to derive the perturbation equations \cite{sahlu2025structure,sahlu2020scalar}, the first order evolution equations derived from eq. (\ref{eq2.47}) are presented as
\begin{eqnarray}
	&&\dot{D}^{m}_{a}=-\gamma Z_{a}+3\Big(\gamma-1\Big)HD^{m}_{a},\label{eq2.48}\\
	&&\dot{Z}_{a}=a\tilde{\bigtriangledown}_{a}\theta+\frac{1-\gamma}{\gamma}\dot{\theta}D^{m}_{a},\label{eq2.49}\\
	&&\dot{\mathcal{C}}_{a}=\mathcal{F}_{a}+\frac{\Big(1-\gamma\Big)\dot{Q}}{\gamma}D^{m}_{a},\label{eq2.50}\\
	&&\dot{\mathcal{F}}_{a}=\frac{\dddot{Q}}{\dot{Q}}\mathcal{C}_{a}+\frac{\Big(1-\gamma\Big)\ddot{Q}}{\gamma}D^{m}_{a}.\label{eq2.51}
\end{eqnarray} Eq. (\ref{eq2.48})--(\ref{eq2.51}) are general evolution equations in the context of $f(Q, \mathcal{L}_{m})$ gravity. In order to investigate the effect of the models A and B on the large scale structure formation, we need to apply each model to these perturbation equations then find the corresponding numerical results.
Using eq. (\ref{eq2.11}), eq. (\ref{eq2.12}) and eq. (\ref{eq2.15}) of the model A, the perturbation equations become
\begin{eqnarray}
	&&\dot{D}^{m}_{a}=-\gamma Z_{a}+3\Big(\gamma-1\Big)HD^{m}_{a},\label{eq2.52}\\
	&&\dot{Z}_{a}=-3\gamma H Z_{a} +\Big(1-\gamma\Big)\Big[\frac{9}{2} H^{2}+\frac{3\beta }{4\alpha}\Big]D^{m}_{a},\label{eq2.53}\\
	&&\dot{\mathcal{C}}_{a}=\mathcal{F}_{a}+\frac{\Big(1-\gamma\Big)\dot{Q}}{\gamma}D^{m}_{a},\label{eq2.54}\\
	&&\dot{\mathcal{F}}_{a}=\frac{\dddot{Q}}{\dot{Q}}\mathcal{C}_{a}+\frac{\Big(1-\gamma\Big)\ddot{Q}}{\gamma}D^{m}_{a},\label{eq2.55}
\end{eqnarray} where we have used
$\dot{\theta}=-\frac{9\gamma H^{2}}{2}+\frac{3\gamma \beta}{4\alpha }$.
Eq.  (\ref{eq2.52})-- (\ref{eq2.55})  are general perturbation equations in the context of $f(Q, \mathcal{L}_{m})$ for model A. Only scalar components of the perturbation equations are believed to enhance most of large scale structure formation mechanism. Using scalar decomposition technique explored in \cite{abebe2023perturbations,abebe2013large,ntahompagaze2022large,munyeshyaka2024covariant} , where $\Delta_{m}=a\tilde{\bigtriangledown}_{a}D^{m}_{a},~ Z=a\tilde{\bigtriangledown}_{a}Z_{a},~ \mathcal{C}=a\tilde{\bigtriangledown}_{a}\mathcal{C}_{a},~ \mathcal{F}=a\tilde{\bigtriangledown}_{a} \mathcal{F}_{a}$  and applying the time derivative on Eq.  (\ref{eq2.52})-- (\ref{eq2.55}), we therefore have the second order ordinary differential equations responsible for structure formation presented as
\begin{eqnarray}
	&&\ddot{\Delta}_{m}=-3H\dot{\Delta}_{m}+\gamma \Big(\gamma-1\Big)\Big[\frac{3\beta}{2\alpha}+9H^{2}\Big]\Delta_{m},\label{eq3.10}\\
	&&\ddot{\mathcal{C}}=\frac{\dddot{Q}}{\dot{Q}}\mathcal{C}+\frac{2\Big(1-\gamma \Big)}{\gamma}\ddot{Q}\Delta_{m}+\dot{Q}\frac{\Big(1-\gamma\Big)}{\gamma}\dot{\Delta}_{m}.\label{eq3.11}
\end{eqnarray}
The energy density resulting from the nonmetricity scalar (eq. (\ref{eq3.11})) couples with the matter energy density. In order to solve eq. (\ref{eq3.10}) and eq. (\ref{eq3.11}), let us first apply the redshift transformation technique considered in different works \cite{sahlu2025structure,sahlu2020scalar,munyeshyaka2024covariant}  presented as $a=\frac{1}{1+z}$, $\dot{f}=-(1+zH)f'$ and $\ddot{f}=\Big(1+z\Big)^{2}H\Big[H'f'+Hf''\Big]$, where $f'$ is the derivative with respect to redshift $z$. The equations in redshift space, when solved, help to compare the results with observations. The resulting equations are presented as 
\begin{figure}
	\includegraphics[width=95mm, height=95mm]{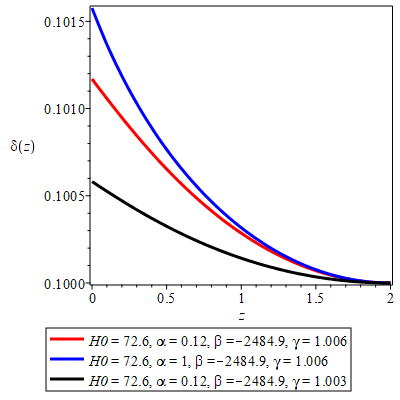}
	\caption{Plot of energy density contrast ($\delta$) versus redshift ($z$) for different values of constant parameters of Eq. (\ref{eq3.12}) and eq. (\ref{eq3.13}) of model A in the context of $f(Q, \mathcal{L}_{m}$ gravity. The choice of values of model parameters was done basing on the constrained parameters obtained in the work done in \cite{hazarika2024f} }
	\label{fig1}
\end{figure}

\begin{figure}
	\includegraphics[width=95mm, height=95mm]{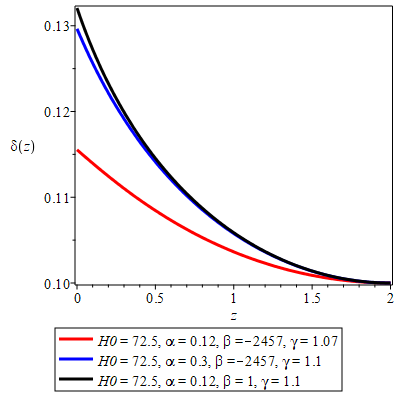}
	\caption{Plot of energy density contrast ($\delta$) versus redshift $z$ for different values of constant parameters of Eq. (\ref{eq3.12}) and eq. (\ref{eq3.13}) of model A. The choice of values of model parameters was done basing on the constrained parameters obtained in the work done in \cite{hazarika2024f} for model B}
	\label{fig2}
\end{figure}
\begin{eqnarray}
	&&\Big(1+z\Big)^{2}H^{2}\Delta''_{m}=-\Big(1+z\Big)H\Big[\Big(1+z\Big)H'-2H\Big]\Delta'_{m}\nonumber\\&&+\gamma \Big(\gamma-1\Big)\Big[\frac{3\beta}{2\alpha}+9 H^{2}\Big]\Delta_{m},\label{eq3.12}\\
	&&\Big(1+z\Big)^{2}H^{2}\mathcal{C}''=-\Big(1+z\Big)H\Big[\Big (1+z\Big) H'+H\Big]\mathcal{C}'+\nonumber\\&&\frac{l_{3}}{l_{1}}   \mathcal{C}+\frac{2\Big( 1-\gamma \Big)}{\gamma}l_{2}\Delta_{m}-l_{1}\frac{\Big(1-\gamma \Big)}{\gamma}\Big(1+z\Big)H\Delta'_{m}\label{eq3.13}
\end{eqnarray}
Eq. (\ref{eq3.12}) and eq. (\ref{eq3.13}) are the energy density perturbation equations responsible for large scale structure formation in the context of $f(Q,\mathcal{L}_{m})$ gravity, where $\Delta''_{m}$ is the second order derivative with respect to redshift of the matter energy density in scalar form ($\Delta_{m}=a\tilde{\bigtriangledown}D^{m}_{a}$), $l_{1}$, $l_{2}$ and $l_{3}$ represent the first, second and third derivatives  with redshift of the nonmetricity scalar. Solving the obtained equations, we need to define the energy density contrast $\delta (z)=\frac{\Delta(z)}{\Delta(z=z_{in})}$ , where $z_{in}$ is the initial redshift. We also define the initial conditions of the energy densities as $\Delta'(z=z_{in}=0$, $\Delta(z=z_{in}=10^{-5}$, $\mathcal{C}'(z=z_{in}=0$, $\mathcal{C}(z=z_{in}=10^{-5}$. By varying the model parameters, numerical results of Eq. (\ref{eq3.12}) and eq. (\ref{eq3.13}) are presented in Fig. (\ref{fig1}) and Fig. (\ref{fig2}). But looking at the figures, the energy density contrast decays as the redshift increases, which implies the perturbation amplitudes are higher in the present universe ($z=0$).

\begin{figure}
	\includegraphics[width=95mm, height=95mm]{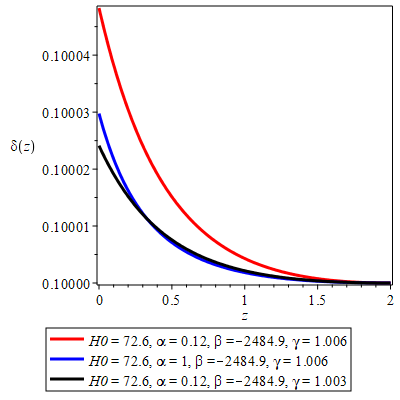}
	\caption{Plot of energy density contrast ($\delta$) versus redshift $z$ for different values of constant parameters of eq. (\ref{eq3.14}) and eq. (\ref{eq3.15}) of model B in the context of $f(Q, \mathcal{L}_{m})$ gravity.}
	\label{fig3}
\end{figure}

\begin{figure}
	\includegraphics[width=95mm, height=95mm]{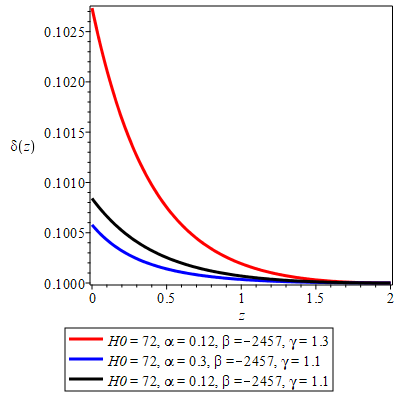}
	\caption{Plot of energy density contrast $\delta$ versus redshift $z$ for different values of constant parameters of eq. (\ref{eq3.14}) and eq. (\ref{eq3.15}) of model B in the context of $f(Q, \mathcal{L}_{m})$ gravity.}
	\label{fig4}
\end{figure}

By going through the same procedures used to obtain the perturbation equations for model A, we can also obtain the perturbation equations for model B. Using Eqs. (\ref{eq2.48})--(\ref{eq2.51}), eqs. (\ref{eq2.30})--(\ref{eq2.32}) together with $\dot{\theta}=-\frac{9\gamma H^{2}}{\gamma+1}-\frac{3\gamma \beta}{2\alpha \Big(\gamma+1\Big)}$, the energy density perturbation equations for model B, in redshift space in the context of $f(G, \mathcal{L}_{m})$ gravity, are given by
\begin{eqnarray}
	&&\Big(1+z\Big)^{2}H^{2}\Delta''_{m}=-\Big(1+z\Big)H\Big[\Big(1+z\Big)H'\nonumber\\&&+H\Big(3\gamma-2-\frac{6\gamma}{\gamma+1}\Big)\Big]\Delta'_{m}+\frac{3 \gamma \Big(1-\gamma \Big)}{\gamma +1}\Big[1+\beta\Big] \Delta_{m},\\\label{eq3.14}
	&&\Big(1+z\Big)^{2}H^{2}\mathcal{C}''=-\Big(1+z\Big)H\Big[\Big (1+z\Big) H'+H\Big]\mathcal{C}'+\nonumber\\&&\frac{l_{3}}{l_{1}} \mathcal{C}
	+\frac{2\Big( 1-\gamma \Big)}{\gamma}l_{2}\Delta_{m}-l_{1}\frac{\Big(1-\gamma \Big)}{\gamma}\Big(1+z\Big)H\Delta'_{m}\label{eq3.15}
\end{eqnarray}
The numerical results of eq. (\ref{eq3.14}) and eq. (\ref{eq3.15}) are presented in Fig. (\ref{fig3}) and Fig. (\ref{fig4}) obtained by interchanging the model parameters. Looking at the plots, the energy density contrast decay with redshift.

After obtaining the energy density perturbation equations and their corresponding numerical results and finding that the energy density contrasts decay with redshift for both models, next step is to investigate the implications of the considered models on the matter power spectrum.

\section{Matter power spectrum in $f(Q, \mathcal{L}_{m})$ gravity}\label{sec4}
In order to get the matter power spectrum for each model, we first put the autonomous dynamical system equations for model A (eq. (\ref{eq2.26})--(\ref{eq2.29})) into eq. (\ref{eq3.12}) and eq. (\ref{eq3.13}) to get the normalised perturbation equations for model A given by
\begin{eqnarray}
	&&\Delta''_{m}=-\frac{1}{\Big(1+z\Big)}\Big[\frac{3\gamma \Big(1+x\Big)}{2}-2\Big]\Delta'_{m}\nonumber\\&&+9\frac{\gamma \Big(\gamma-1\Big)}{\Big(1+z\Big)^{2}} \Big(x+1\Big)\Delta_{m},\label{eq4.1}\\
	&&\mathcal{C}''=-\frac{1}{\Big(1+z\Big)}\Big[\frac{3\gamma \Big(1+x\Big)}{2}+1\Big]\mathcal{C}'+\frac{\beta l_{3}}{6\Big(1+z\Big)^{2}\alpha x l_{1}}\mathcal{C}\nonumber\\&&
	+\frac{\beta \Big( 1-\gamma \Big)}{3\Big(1+z\Big)^{2}\alpha x \gamma}l_{2}\Delta_{m}-l_{1}\frac{\Big(1-\gamma \Big)}{\gamma\Big(1+z\Big)}\Big(\frac{\beta}{6\alpha x}\Big)^{\frac{1}{2}}\Delta'_{m}\label{eq4.2}
\end{eqnarray}
\begin{figure}
	\includegraphics[width=95mm, height=95mm]{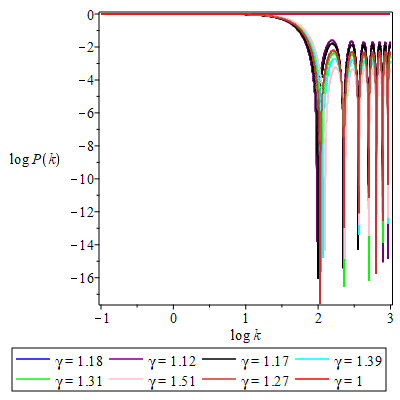}
	\caption{Plot matter power spectra for different values of $\gamma$ for $\alpha=0.3$ and $\beta=-0.0001$ of eq. (\ref{eq4.1}) and eq. (\ref{eq4.2}) of model A in the context of $f(Q, \mathcal{L}_{m})$ gravity. We use the initial conditions $\Delta(z)=10^{-5}$, $\Delta'(z)=0$, $\mathcal{C}(z)=10^{-5}$, $\mathcal{C}'(z)=0$.}
	\label{fig5}
\end{figure}
\begin{figure}
	\includegraphics[width=95mm, height=95mm]{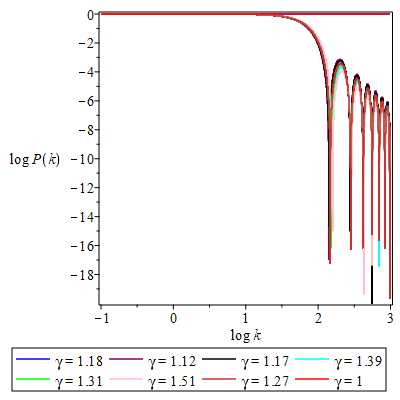}
	\caption{Plot matter power spectra for different values of $\gamma$ for $\alpha=0.3$ and $\beta=-0.0001$ of eq. (\ref{eq4.1}) and eq. (\ref{eq4.2}) of model A in the context of $f(Q, \mathcal{L}_{m})$ gravity. We use the initial conditions $\Delta(z)=10^{-5}$, $\Delta'(z)=10^{-3}$, $\mathcal{C}(z)=10^{-5}$, $\mathcal{C}'(z)=10^{-3}$.}
	\label{fig6}
\end{figure}

\begin{figure}
	\includegraphics[width=95mm, height=95mm]{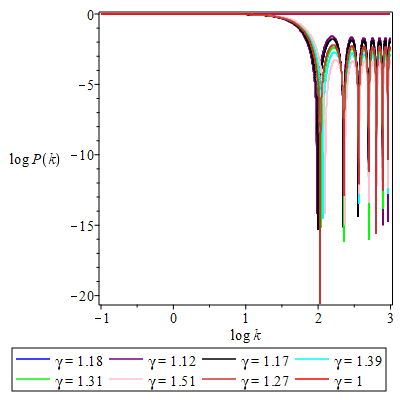}
	\caption{Plot matter power spectra for different values of $\gamma$ for $\alpha=0.3$ and $\beta=-0.0001$ of eq. (\ref{eq4.1}) and eq. (\ref{eq4.2}) of model A in the context of $f(Q, \mathcal{L}_{m})$ gravity. We use the initial conditions $\Delta(z)=10^{-5}$, $\Delta'(z)=10^{-8}$, $\mathcal{C}(z)=10^{-5}$, $\mathcal{C}'(z)=10^{-8}$.}
	\label{fig7}
\end{figure}
\begin{figure}
	\includegraphics[width=95mm, height=95mm]{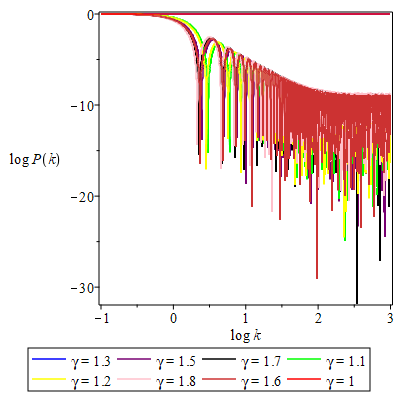}
	\caption{Plot matter power spectra for different values of $\gamma$ for $\alpha=0.4$ and $\beta=-0.0001$ of eq. (\ref{eq4.3}) and eq. (\ref{eq4.4}) of model B in the context of $f(Q, \mathcal{L}_{m})$ gravity. We use the initial conditions $\Delta(z)=10^{-5}$, $\Delta'(z)=10^{-3}$, $\mathcal{C}(z)=10^{-5}$, $\mathcal{C}'(z)=10^{-3}$.}
	\label{fig8}
\end{figure}

By assuming an isotropic FRW universe, the matter power spectrum in the context of $f(Q,\mathcal{L}_{m})$ gravity, which can be compared to the $\Lambda$CDM predictions at any given scale $k$ can be represented as  \cite{abebe2013large,ntahompagaze2022large, munyeshyaka1}
\begin{equation*}
	P^{f(Q,\mathcal{L}_{m}})_{k}=(|\Delta(k)|)^{2}.
\end{equation*}
Put the autonomous dynamical system equations for model B (eq. (\ref{eq2.43})--(\ref{eq2.46})) into the perturbation equations eq. (\ref{eq3.14}) and eq. (\ref{eq3.15}) to get normalised perturbation equations for model B presented as 

\begin{eqnarray}
	&&\Delta''_{m}=-\frac{1}{\Big(1+z\Big)}\Big[\frac{3 \Big(1+y\Big)}{1+\gamma}+\Big(3\gamma-2-\frac{6\gamma}{\gamma+1}\Big)\Big]\Delta'_{m}\nonumber\\&&+\frac{3 \gamma \Big(1-\gamma \Big)}{\gamma +1}\Big[1+\beta\Big]\Delta_{m},\label{eq4.3}\\
	&&\mathcal{C}''=-\frac{1}{\Big(1+z\Big)}\Big[\frac{3 \Big(1+y\Big)}{1+\gamma}+1\Big]\mathcal{C}'+\frac{\beta l_{3}}{6\Big(1+z\Big)^{2}\alpha y l_{1}}   \mathcal{C}
	\nonumber\\&& +\frac{\beta \Big( 1-\gamma \Big)}{3\Big(1+z\Big)^{2}\alpha y \gamma}l_{2}\Delta_{m}-l_{1}\frac{\Big(1-\gamma \Big)}{\gamma\Big(1+z\Big)}\Big(\frac{\beta}{6\alpha y}\Big)^{\frac{1}{2}}\Delta'_{m}\label{eq4.4}
\end{eqnarray}
In order to compute the matter spectra for each model, we consider 3 different sets of initial conditions for the system of equations eq. (\ref{eq4.1}) and eq. (\ref{eq4.2}) for model A and eq. (\ref{eq4.3}) and eq. (\ref{eq4.4}) for model B, in order to interpret the sensitivity of the models on the matter spectra. To obtain the numerical results,for model A, we solve simultaneously eq. (\ref{eq2.26})--eq. (\ref{eq2.29}) to get redshift-dependent solutions for parameters $x$, $\Omega_{1}$ and $\Omega_{2}$. We then use the obtained solutions together with the solutions of eq. (\ref{eq4.1}) and eq. (\ref{eq4.2}) to get the matter power spectra for model A. The power spectra for model A are presented in Fig. (\ref{fig5}) and Fig. (\ref{fig6}).  For model B, we solve eq. (\ref{eq2.43})--eq. (\ref{eq2.46}) to obtain redshift-dependent solutions for the parameters $y$, $\Omega_{3}$ and $\Omega_{4}$. Using these solutions together with the solutions of eq. (\ref{eq4.3}) and eq. (\ref{eq4.4}), we compute the matter power spectra for model B. The power spectra for model B are presented in Fig. (\ref{fig7}) and Fig. (\ref{fig8}).
\begin{figure}
	\includegraphics[width=95mm, height=95mm]{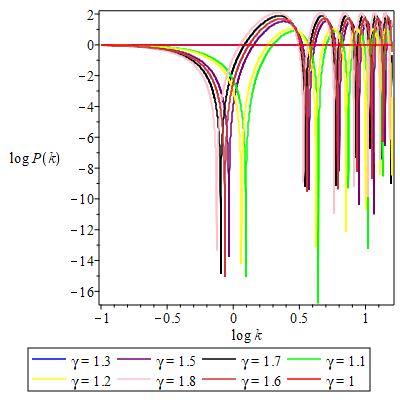}
	\caption{Plot matter power spectra for different values of $\gamma$ for $\alpha=0.4$ and $\beta=-0.0001$ of eq. (\ref{eq4.3}) and eq. (\ref{eq4.4}) of model B in the context of $f(Q, \mathcal{L}_{m})$ gravity. We use the initial conditions $\Delta(z)=10^{-5}$, $\Delta'(z)=10^{-8}$, $\mathcal{C}(z)=10^{-5}$, $\mathcal{C}'(z)=10^{-8}$.}
	\label{fig9}
\end{figure}
\begin{figure}
	\includegraphics[width=95mm, height=95mm]{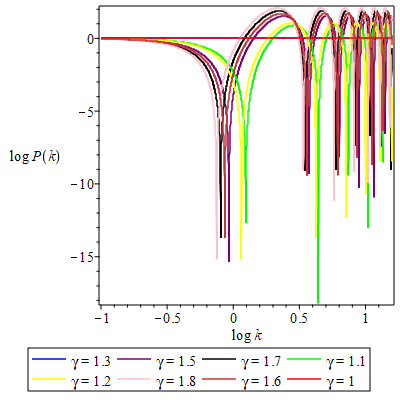}
	\caption{Plot matter power spectra for different values of $\gamma$ for $\alpha=0.4$ and $\beta=-0.0001$ of eq. (\ref{eq4.3}) and eq. (\ref{eq4.4}) of model B in the context of $f(Q, \mathcal{L}_{m})$ gravity. We use the initial conditions $\Delta(z)=10^{-5}$, $\Delta'(z)=0$, $\mathcal{C}(z)=10^{-5}$, $\mathcal{C}'(z)=0$.}
	\label{fig10}
\end{figure}
\newpage

\section{Structure growth and data analysis}\label{sec5}
In this section, it is worth to extend the analysis of the perturbation equations and its role on structure formation to observational context. In doing so, we first derive a structure growth equation, then use different observational data sets to discuss the compatibility of the models with observation. Let us first seek the growth equation in the following subsection.
\subsection{Structure growth equation in the context of $f(Q, \mathcal{L}_{m})$ gravity}
To constrain the matter energy density fluctuations with observations, we consider the redshift-space distortions (RSD) discussed mainly in \cite{kazantzidis2018evolution,kazantzidis2021hints,kazantzidis2021sigma,panotopoulos2021growth}. Following these works, we compute the combinations $f\sigma 8$ in the context of $f(Q, \mathcal{L}_{m})$ gravity. We first present the linear growth rate $f$ as 
\begin{equation}
	f=-(1+z)\frac{\Delta'_{m}(z)}{\Delta_{m}(z_{in})}, \label{eq39}
\end{equation}
and using eq. (\ref{eq39}), together with the quasi-static approximation \cite{sahlu2025structure}, eqs. (\ref{eq3.12}) and (\ref{eq3.13}) can be related to the linear growth rate as 
\begin{eqnarray}
	&&\Big(1+z\Big)f'=-f^{2}-\Big[(1+z)\frac{E'}{E}-3\Big]f-9\gamma \Big(\gamma-1\Big)-\frac{3 \gamma (\gamma-1)\beta}{2\alpha H^{2}_{0} E^{2}},\label{eq40}
\end{eqnarray} where $E^{2}=\frac{H^{2}}{H^{2}_{0}}$ is obtained from eq. (\ref{eq2.11}) as
\begin{eqnarray}
	E^{2}-\frac{1}{\alpha}\Big[\Omega_{m}(1+z)^{3}+\frac{\beta}{12H^{2}_{0}}\Big]=\Omega_{m}\Big(1+z\Big)^{3}.
\end{eqnarray}
Solving eq. (\ref{eq40}) numerically for $f(z)$ and combining $f(z)$ with $\sigma 8$, where $\sigma 8$ is the root mean square normalisation of the matter power spectrum within the radius sphere of $8h^{-1} MPC$ \citep{sahlu2025structure} defined as
\begin{eqnarray}
	\sigma 8(z)=\sigma 8(z_{in})\frac{\Delta_{m}(z)}{\Delta_{m}(z_{in})},
\end{eqnarray}  the combination $f\sigma 8(z)$ can be presented as \citep{kazantzidis2018evolution,kazantzidis2021sigma,nesseris2017tension,sahlu2025structure,panotopoulos2021growth}
\begin{eqnarray}
	f\sigma 8 (z)=-(1+z)\sigma 8 (z)\frac{\Delta'_{m}(z)}{\Delta_{m}(z_{in})}. \label{eq41}
\end{eqnarray}
Eq. (\ref{eq40}) and Eq. (\ref{eq41}) will be used to confront the model parameters with the RSD data set using MCMC. The values $\delta'_{m}(z)$, $\delta_{m}(z)$ and $\delta_{m}(z_{in})$ are obtained  by solving numerically eqs. (\ref{eq3.12}) and (\ref{eq3.13})  for a given set of initial conditions. This theoretical results can now be used  to constrain the parameters ($\Omega_{m}$, $\sigma~8$,$\gamma$, $H_{0}$, $\alpha$ and $\beta$) using $f\sigma~8$ data in the context of $f(G)$ gravity model. In the next section, we introduce different observational data sets to be considered to compare our $f(Q,\mathbf{L}_{m})$ gravity model with observation using MCMC.
\section{Data analysis and Results}\label{sec:resutls}
In this section, we discuss and present the observational data used to constrain the model parameters using the MCMC analysis.
\subsection{Data analysis}
\begin{itemize}
	\item \textbf{CC Data:} This data set contains 32 model-independent observational points corresponding to the Hubble parameter, commonly referred to as Cosmic Chronometers (CC) \cite{Jimenez:2001gg, Moresco:2015cya}.  The covariance matrix constructed in \cite{Moresco:2012jh, Moresco:2015cya, Moresco:2016mzx} is used to evaluate the likelihood.\footnote{The Python code to construct the covariance matrix for the $15$ highly correlated samples is available at \url{https://gitlab.com/mmoresco/CCcovariance}} 
	\item \textbf{PP Data:} This data set consists of 1550 spectroscopically confirmed Type Ia supernovae \cite{Brout:2022vxf}. The full catalog contains 1770 data samples, out of which we use 1590 samples by applying filter to the redshift $z>0.01$. The non-Cepheid calibrated samples are used corresponding to the observational column apparent magnitude \(m_{\rm obs}\). The nuisance parameter $M_B$ has been analytically marginalized. We refer to this data set as `PP'.\footnote{We analytically marginalize $M_B$ following the prescription in \cite{Goliath:2001af}, which is also implemented in the \texttt{Cobaya} repository {Likelihood estimation code: \url{https://github.com/CobayaSampler/cobaya/blob/master/cobaya/likelihoods/sn/pantheonplus.py}}}

	\item \textbf{DES Data:} This data set contains Type Ia Supernovae samples from a different observation -- the Dark Energy Survey (DES-SN5YR) -- which includes 1829 distinct SNe \cite{DES:2024jxu,DES:2024hip}. It consists of 194 nearby SNe samples with redshift \(0<z < 0.1\) and 1635 DES SNe samples. The catalog provides the distance modulus \(\mu\) along with the full covariance matrix. The distance modulus is defined as:
	\begin{equation}
		\mu \equiv m-M_b = 5\log(D_L/\text{Mpc}) + 25 \  ,
	\end{equation}
	where, \(m\) denotes the apparent magnitude of the supernova, \(M_b\) is the absolute magnitude and \(D_L\) is the luminosity distance: 	
	\begin{equation}
		D_L({z}) = c(1+{z}) \int_0^{{z}} \frac{dz'}{H(z')} \ ,
	\end{equation} 
	assuming a flat FLRW metric, and \(c\) is the speed of light in km/s. The model parameters are constrained by minimizing the chi-square ($\chi^2$) likelihood, defined as:
	\begin{equation}
		-2 \ln (\mathcal{L}) = \chi^2 = \rm \Delta D_{i} \mathcal{C}^{-1}_{ij} \Delta D_j\ ,
	\end{equation}
	where $\rm	\Delta D = \mu_{\rm Obs} - \mu_{\rm Model}  .$ The likelihood is computed by marginalizing $M_B$ using the code available in the \href{https://github.com/des-science/DES-SN5YR/blob/main/5_COSMOLOGY/SN_only_cosmosis_likelihood.py}{DES-SN5YR module}. In this analysis, PP and DES samples are not combined, as both catalogs share some overlapping data points. We label this data set as `DES'.

	\item \textbf{DESI BAO:} This data set includes samples of Baryon Acoustic Oscillations (BAO) from the Dark Energy Spectroscopic Instrument (DESI) Release II \cite{DESI:2025zgx}. The observables are \(\{D_M/r_d, D_H/r_d, D_V/r_d\}\), where \(D_M\) denotes the comoving angular diameter distance, \(D_H\) the Hubble distance, \(D_V\) the spherically averaged distance, and \(r_d\) the sound horizon at the drag epoch, corresponding to the redshift \(z_d = 1060.0\) \cite{eBOSS:2020yzd,DESI:2024mwx,Planck:2018vyg},
	\begin{equation}
		r_{d} =\int_{ z_{d}}^{\infty} \dfrac{3 \times 10^{5} d{z}}{H \sqrt{3\left(1 + \frac{3 \Omega_{b_0}h^2}{4 \Omega_{\gamma_0}h^2(1+{z})}\right)}} \ .
		\label{sound_distance}
	\end{equation}
	Here,  \(\Omega_{\gamma_0} h^2\) denotes the photon density parameter and \(h \equiv H_0/100\), with the value \(2.472 \times 10^{-5}\) and $\Omega_{b_0}$ represents the baryon density at the present epoch with the value $\Omega_{b_0}h^2 = 0.02236$ \cite{Planck:2018vyg,Chen:2018dbv}. However, for this study we will treat \(r_d\) as a free parameter. We refer to this data set as `BAO'.
	
	\item \textbf{\boldmath $f\sigma_8$ Data :} This data sets includes $30$ observational samples of redshift-space distortion $f\sigma_8$ in the redshift \(z \in [0.001,1.944]\) \cite{sahlu2025structure}. We label this data set as `fs8'. 
	
\end{itemize}
For the above data sets, we compute the joint likelihood corresponding to following combinations of datasets: (i) CC+PP (ii) CC+DES (iii) CC+BAO+fs8 (iv) CC+BAO+fs8+PP (v) CC+BAO+fs8+DES. The joined likelihood is estimated as
\begin{equation}
	-2 \ln \mathcal{L}_{\rm tot} = \chi^2_{\rm tot}.
\end{equation}
The likelihood is estimated by implementing the model in \texttt{Python} using the publicly available affine-invariant Markov Chain Monte Carlo (MCMC) ensemble sampler \texttt{emcee} \cite{Foreman-Mackey:2012any}. The resulting posterior distributions are visualized using triangular (corner) plots, generated by analyzing the MCMC chains with \texttt{GetDist} \cite{Lewis:2019xzd}.

\subsection{Results}

\begin{table}
	\centering
	\begin{tabular}{l r }
		\hline
		\hline
		\multicolumn{2}{c}{\ttfamily Model A}  \\
		\hline
		Parameters & Range \\
		\hline
		$H_0$ & $[30,100]$ \\
		$\alpha$ & [0,0.5]   \\
		$\beta$ & $ [-4000, -1000]$ \\
		$\gamma$ & [1.0, 2.0] \\
		$r_d$  & [100, 300] \\
		\(\sigma_8\) & [0, 0.9] \\

		\hline
		\hline
	\end{tabular}
	\caption{The uniform prior range on the model parameters.}
	\label{tab:prior_range}
\end{table} 

In this section, we present the parameter estimation for Model A by varying the free parameters associated with the Hubble parameter (see Eq.~\eqref{eq2.15}), alongside other key cosmological quantities—namely, the absolute magnitude of Type Ia supernovae $M_b$, the sound horizon at the end of the drag epoch $r_d$, and the root-mean-square normalization of the matter power spectrum $\sigma_8$, defined within a sphere of radius $8h^{-1}$ Mpc. A uniform prior is adopted for each parameter as summarized in Tab.~(\ref{tab:prior_range}).

\begin{figure}
	\centering
	\includegraphics[scale=0.5]{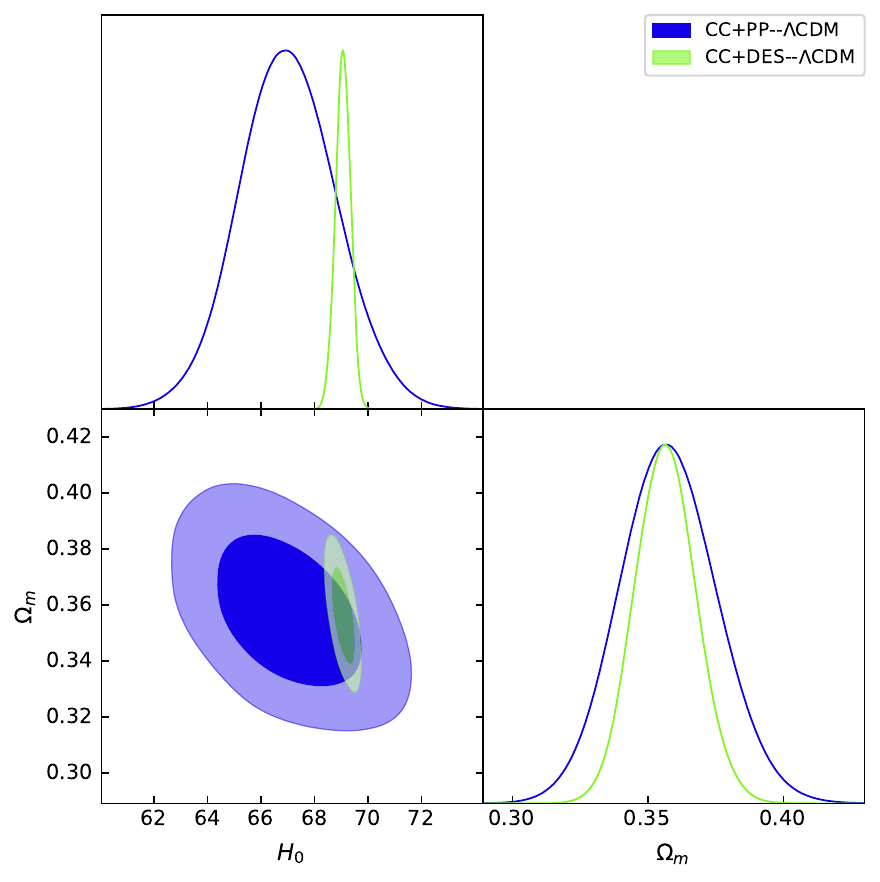}
	\includegraphics[scale=0.5]{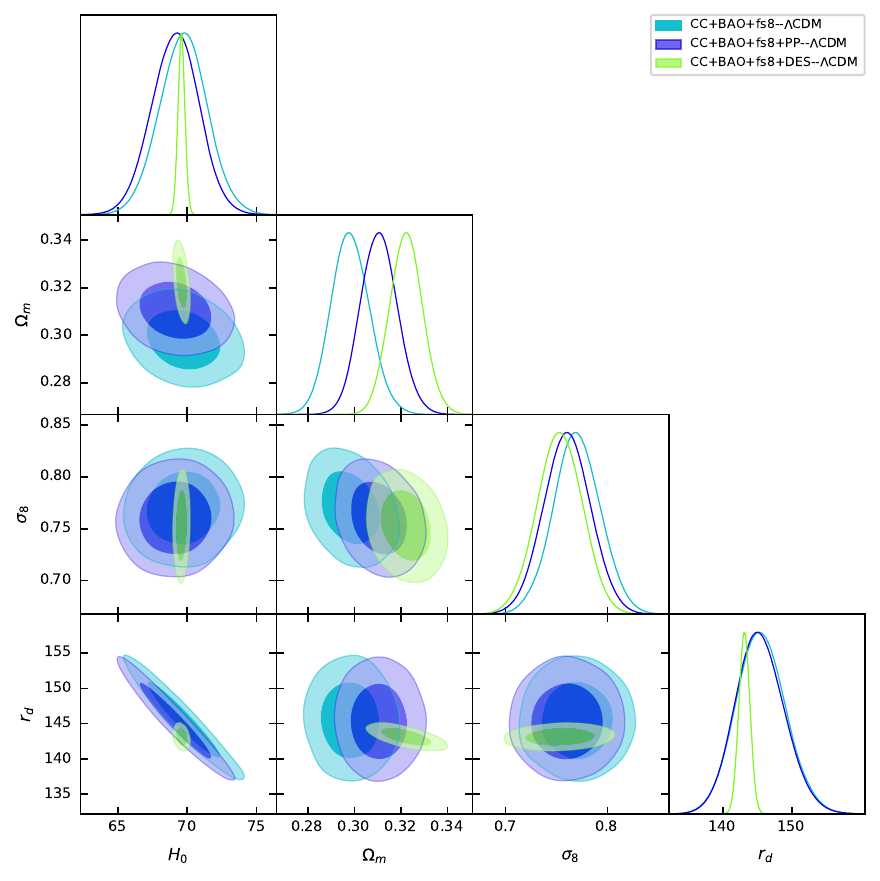}
	\caption{The 1D 2D marginalized parameter distribution for \(\Lambda\)CDM marginalizing $M_b$.  }
	\label{fig:lcdm_post}
\end{figure}

\begin{figure}
	\centering
	\includegraphics[scale=0.5]{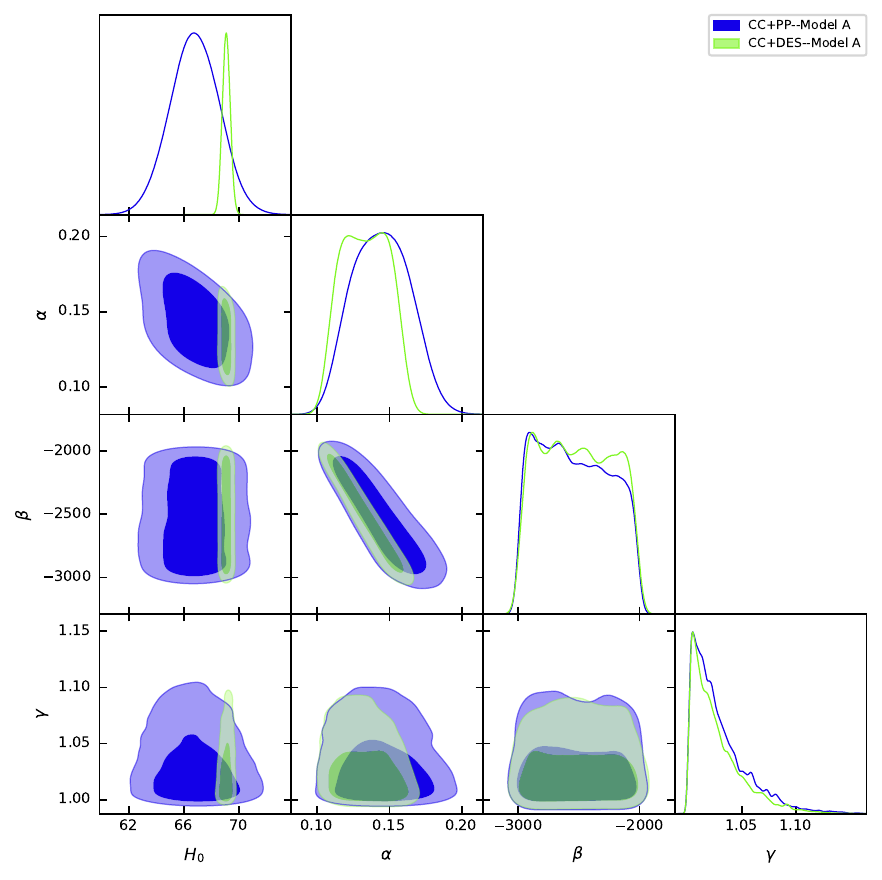}
	\includegraphics[scale=0.5]{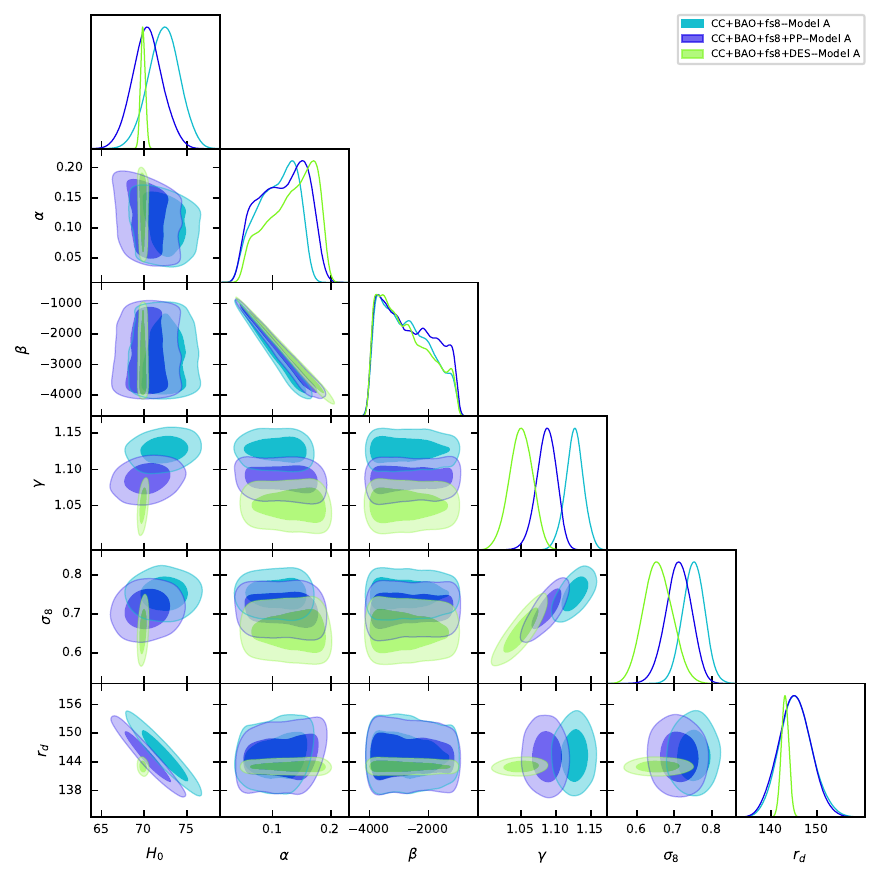}
	\caption{The posterior distribution for Model A marginalizing $M_b$.}
	\label{fig:corner_mod1_sn}
\end{figure}

We begin by testing the flat $\Lambda$CDM model as a baseline, varying $H_0$ and $\Omega_{m_0}$ over the ranges $[30,100]$ and $[0,0.8]$, respectively. The resulting posterior distributions, marginalized over $M_b$, are shown in Fig.~(\ref{fig:lcdm_post}), and the best-fit values are presented in Tabs.~(\ref{tab:modelA_sn_params}) and~(\ref{tab:modelA_fs8_params}).\\ 
Model A is tested using two separate Type Ia supernova catalogs, each yielding distinct values of $H_0$ as seen in Tab.~(\ref{tab:modelA_sn_params}), indicating tension between the datasets. However, the other cosmological parameters remain in good agreement. The recovered values of $H_0$ are consistent with those from $\Lambda$CDM, suggesting that the discrepancy arises primarily due to differences in calibration between the datasets.\\ 
\begin{table}
	\centering
	\begin{tabular} { l  c c}
		\noalign{\vskip 3pt}\hline\noalign{\vskip 1.5pt}\hline\noalign{\vskip 5pt}
		\multicolumn{1}{c}{\bf } &  \multicolumn{1}{c}{\bf CC+PP--$\Lambda$CDM} &  \multicolumn{1}{c}{\bf CC+DES--$\Lambda$CDM}\\
		\noalign{\vskip 3pt}\cline{2-3}\noalign{\vskip 3pt}
		
		Parameter &  68\% limits &  68\% limits\\
		\hline
		{\boldmath$H_0            $} & $68.2\pm 2.5               $ & $69.05\pm 0.28             $\\
		
		{\boldmath$\Omega_m       $} & $0.330\pm 0.018            $ & $0.358\pm 0.012            $\\
		
		\hline
		\hline
	\end{tabular}
	\begin{tabular} { l  c c}
		\noalign{\vskip 3pt}\hline\noalign{\vskip 1.5pt}\hline\noalign{\vskip 5pt}
		\multicolumn{1}{c}{\bf } &  \multicolumn{1}{c}{\bf CC+PP--Model A} &  \multicolumn{1}{c}{\bf CC+DES--Model A}\\
		\noalign{\vskip 3pt}\cline{2-3}\noalign{\vskip 3pt}
		
		Parameter &  68\% limits &  68\% limits\\
		\hline
		{\boldmath$H_0            $} & $67.6\pm 2.5               $ & $69.10\pm 0.29             $\\
		
		{\boldmath$\alpha         $} & $0.135\pm 0.019            $ & $0.135^{+0.019}_{-0.015}   $\\
		
		{\boldmath$\beta          $} & $-2550^{+200}_{-410}       $ & $-2540^{+210}_{-420}       $\\
		
		{\boldmath$\gamma         $} & $1.047^{+0.014}_{-0.046}   $ & $1.0271^{+0.0062}_{-0.027} $\\

		\hline
		\hline
	\end{tabular}
	\caption{The model A parameter's best fit values obtained from MCMC. }
	\label{tab:modelA_sn_params}
\end{table}
The Pantheon+ (PP) dataset can not constraint $H_0$ as it exhibits strong degeneracy with $M_b$. To break this, we include external Cosmic Chronometer (CC) data, resulting in a value of $H_0 \approx 68.2$ km/s/Mpc—a mildly higher from Planck Collaboration's result~\cite{Planck:2018vyg}. In contrast, the DES catalog, leading to $H_0 \approx 69.05$ km/s/Mpc—closer to the SH0ES result~\cite{Brout:2022vxf}.\\ 
We further extend the analysis by including BAO and $f\sigma_8$ datasets and re-evaluating both Model A and $\Lambda$CDM. The corresponding best-fit values are listed in Tab.~(\ref{tab:modelA_fs8_params}), and the posterior distributions are shown in Fig.~(\ref{fig:lcdm_post}). The inclusion of these datasets brings all three combinations to a consistent $H_0 \sim 69.8$ km/s/Mpc. The values of $\sigma_8 \sim 0.769$ and $r_d \sim 144.5$ Mpc across datasets for flat $\Lambda$CDM. \\ 
When Model A is tested with the $f\sigma_8$ datasets, the posterior distributions shown in Fig.~(\ref{fig:corner_mod1_sn}) and the best-fit parameters in Tab.~(\ref{tab:modelA_fs8_params}) indicate $H_0 \sim 70.0$ km/s/Mpc for both CC+BAO+fs8+PP and CC+BAO+fs8+DES combinations—slightly higher than the corresponding $\Lambda$CDM result. The effective equation of state parameter $\gamma$ for the background fluid is found to be $\gamma \sim 1.07$ for both combinations, indicating a mild deviation from pressure-less dark matter ($\gamma = 1$). A similar value is also obtained for CC+PP and CC+DES data sets ($\gamma \sim 1.03$). The derived value of $r_d$  remain consistent with $\Lambda$CDM, though $\sigma_8$ shows noticeable deviation. 

Specifically, the DES dataset yields $\sigma_8 \sim 0.658$ for Model A, which deviates from $\Lambda$CDM by about $2.14\sigma$. This suggests a potential resolution to the $\sigma_8$ tension in this context.  Using the CC+BAO+fs8 combination alone, Model A yields $H_0 \sim 72.2$ km/s/Mpc—very close to the SH0ES value. However, in this case, $\gamma \sim 1.127$, indicating a significant deviation from a pressure-less fluid. Despite this, $\sigma_8$ and $r_d$ remain compatible with $\Lambda$CDM values. In addition to that a mild but significant up lift can be noticed in the value of \(H_0 \sim 70.1\) km/s/Mpc for CC+BAO+fs8+PP and CC+BAO+fs8+DES compare to fiducial model.

In summary, while Model A can yield a higher $H_0$, consistent with local measurements, the inclusion of larger and more diverse datasets tends to pull the model closer to the $\Lambda$CDM values, limiting its ability to fully resolve the Hubble tension. Nonetheless, it offers a compelling and physically viable alternative to the standard cosmological model.
\begin{table*}
	\caption{The parameter's best fit values obtained from MCMC. The dashed shows that the corresponding parameter is not fitted for that particular data set. }
		\begin{tabular} { l  c c c}
			\noalign{\vskip 3pt}\hline\noalign{\vskip 1.5pt}\hline\noalign{\vskip 5pt}
			\multicolumn{1}{c}{\bf } &  \multicolumn{1}{c}{\bf CC+BAO+fs8--$\Lambda$CDM} &  \multicolumn{1}{c}{\bf CC+BAO+fs8+PP--$\Lambda$CDM} &  \multicolumn{1}{c}{\bf CC+BAO+fs8+DES--$\Lambda$CDM}\\
			\noalign{\vskip 3pt}\cline{2-4}\noalign{\vskip 3pt}
			
			Parameter &  68\% limits &  68\% limits &  68\% limits\\
			\hline
			{\boldmath$H_0            $} & $70.1\pm 2.4               $ & $69.5\pm 2.4               $ & $69.62\pm 0.25             $\\
			
			{\boldmath$\Omega_m       $} & $0.2968\pm 0.0083          $ & $0.3030\pm 0.0078          $ & $0.3211\pm 0.0071          $\\
			
			{\boldmath$\sigma_8       $} & $0.771\pm 0.023            $ & $0.766\pm 0.023            $ & $0.752\pm 0.022            $\\
			
			{\boldmath$r_d            $} & $144.9^{+4.5}_{-5.3}       $ & $145.5\pm 4.9              $ & $143.13\pm 0.80            $\\
			\hline
			\hline
		\end{tabular}
		\begin{tabular} { l  c c c}
			\noalign{\vskip 3pt}\hline\noalign{\vskip 1.5pt}\hline\noalign{\vskip 5pt}
			\multicolumn{1}{c}{\bf } &  \multicolumn{1}{c}{\bf CC+BAO+fs8--Model A} &  \multicolumn{1}{c}{\bf CC+BAO+fs8+PP--Model A} &  \multicolumn{1}{c}{\bf CC+BAO+fs8+DES--Model A}\\
			\noalign{\vskip 3pt}\cline{2-4}\noalign{\vskip 3pt}
			
			Parameter &  68\% limits &  68\% limits &  68\% limits\\
			\hline
			{\boldmath$H_0            $} & $72.2\pm 2.5               $ & $70.6^{+2.3}_{-2.6}        $ & $69.88\pm 0.28             $\\
			
			{\boldmath$\alpha         $} & $0.100\pm 0.013            $ & $0.109\pm 0.015            $ & $0.120^{+0.016}_{-0.014}   $\\
			
			{\boldmath$\beta          $} & $-2540^{+250}_{-420}       $ & $-2520^{+300}_{-420}       $ & $-2530^{+210}_{-410}       $\\
			
			{\boldmath$\gamma         $} & $1.127\pm 0.012            $ & $1.098^{+0.014}_{-0.012}   $ & $1.051^{+0.017}_{-0.015}   $\\
			
			{\boldmath$\sigma_8       $} & $0.752\pm 0.029            $ & $0.725\pm 0.033            $ & $0.658\pm 0.038            $\\
			
			{\boldmath$r_d            $} & $145.7\pm 4.9              $ & $145.7\pm 4.9              $ & $142.94\pm 0.83            $\\
			\hline
			\hline
			\label{tab:modelA_fs8_params}
		\end{tabular}
\end{table*}

\section{Discussions and Conclusion}\label{sec7}
In this study, we explored the energy density perturbations and matter power spectrum within the framework of $f(Q,\mathcal{L}_{m})$ gravity, where $Q$ represents the nonmetricity scalar and $\mathcal{L}_{m}$ denotes the matter Lagrangian.  We used two different cosmological models namely model A and model B defined as $f=-\alpha Q+2l_{m}+\beta$ and $f=-\alpha Q+(2l_{m})^{2}+\beta$ respectively,  to analyse their effects on perturbation levels and on the matter power spectrum in the context of $f(Q,\mathcal{L}_{m})$ gravity, where $\alpha$ and $\beta$ are model parameters. Using dynamical system analysis, we obtain the autonomous ordinary differential equations for each models, namely eq. (\ref{eq2.26})--(\ref{eq2.29}) and eq. (\ref{eq2.43})--(\ref{eq2.46}). These autonomous equations are helpful in computing matter power spectrum. Defining gradient variables responsible for large scale structure formation and using  scalar decomposition, harmonic decomposition together with redshift transformation techniques, we derive  and present the energy density perturbation equation in redshift space (eq. (\ref{eq3.12})--(\ref{eq3.13}) for model A and eq. (\ref{eq3.14})--(\ref{eq3.15}) for model B) using the $1+3$ covariant formalism. These perturbation equations are believe to enhance matter clustering, hence responsible for structure formation. Using different initial conditions such as $\Delta(z_{in})=10^{-5}$, $\Delta'(z_{in})=0$, $\mathcal{C}(z_{in})=10^{-5}$ and $\mathcal{C}'(z_{in})=0$ and defining the energy density contrast as $\delta(z)=\frac{\Delta(z)}{\Delta(z_{in})}$, we solve numerically the perturbation equations for both models and the results are presented in Fig. (\ref{fig1}) and Fig. (\ref{fig2}) for model A, whereas Fig. (\ref{fig3}) and Fig. (\ref{fig4}) represent numerical results for model B. For model A, the energy density contrast decays with redshift. By changing the model parameters such as $\alpha$ and $\gamma$, the amplitudes of the $\delta(z)$ change significantly as can be shown in Fig. (\ref{fig1}) and Fig. (\ref{fig2}). By looking at Fig. (\ref{fig3}) and Fig. (\ref{fig4})  of model B, the energy density contrast ($\delta(z)$) decays with redshift and the amplitudes are affected by changing model parameters.  For both models, the $\delta(z)$ decays with redshift but dies off quickly for model B than model A as the redshift increases. After solving numerically the perturbation equations for both models and obtaining the density contrast, we use the obtained autonomous ordinary differential equations (eq. (\ref{eq2.26})--(\ref{eq2.29})) together with the perturbation equations eq. (\ref{eq3.12})--(\ref{eq3.13} for model A to compute matter power spectra in the context of $f(Q,\mathcal{L}_{m})$ gravity. The matter power spectra are presented in Fig. (\ref{fig5}) using $\Delta(z_{in})=10^{-5}$, $\Delta'(z_{in})=0$, $\mathcal{C}(z_{in})=10^{-5}$ and $\mathcal{C}'(z_{in})=0$ as initial conditions, in Fig. (\ref{fig6}) using $\Delta(z_{in})=10^{-5}$, $\Delta'(z_{in})=10^{-3}$, $\mathcal{C}(z_{in})=10^{-5}$ and $\mathcal{C}'(z_{in})=10^{-3}$ as initial conditions and in Fig. (\ref{fig7}) using $\Delta(z_{in})=10^{-5}$, $\Delta'(z_{in})=10^{-8}$, $\mathcal{C}(z_{in})=10^{-5}$ and $\mathcal{C}'(z_{in})=10^{-8}$ as initial conditions. Furthermore, using the obtained autonomous ordinary differential equations (eq. (\ref{eq2.43})--(\ref{eq2.46})) together with the perturbation equations eq. (\ref{eq3.14})--(\ref{eq3.15} for model B, we compute matter power spectra in the context of $f(Q,\mathcal{L}_{m})$ gravity. The matter power spectra are presented in Fig. (\ref{fig8}) using $\Delta(z_{in})=10^{-5}$, $\Delta'(z_{in})=10^{-3}$, $\mathcal{C}(z_{in})=10^{-5}$ and $\mathcal{C}'(z_{in})=10^{-3}$ as initial conditions, in Fig. (\ref{fig9}) using $\Delta(z_{in})=10^{-5}$, $\Delta'(z_{in})=10^{-8}$, $\mathcal{C}(z_{in})=10^{-5}$ and $\mathcal{C}'(z_{in})=10^{-8}$ as initial conditions and in Fig. (\ref{fig10}) using $\Delta(z_{in})=10^{-5}$, $\Delta'(z_{in})=0$, $\mathcal{C}(z_{in})=10^{-5}$ and $\mathcal{C}'(z_{in})=0$ as initial conditions. By looking at all matter power spectra for both models, the curves decay and remain under the General Relativity (GR) scale invariant line as $k$ increase for model A, whereas the curves decay and evolve above the GR invariant line for model B. As the initial conditions changes, there appear a slight changes in the behaviour of the matter power spectra curves. In the present work, one can highlight that 
\begin{itemize}
	\item The obtained matter contrasts $\Delta(z)$ couple with the the density contrast resulting from the $f(Q,\mathcal{L}_{m})$ models $\mathcal{C}(z)$, therefore  $\mathcal{C}(z)$ influences the behavior of both energy density contrasts and matter power spectra for both models.
	\item The obtained density contrasts $\delta(z)$ decay with redshift for both models
	\item The obtained matter power spectra decay with a change in amplitude and remain below GR invariant line for model A and evolve above GR invariant line for model B as $k$ increases for different values of $\gamma$ and initial conditions.
	\item For all sets of initial conditions, there was no oscillations behaviour pictured as identified in the work carried out by\cite{fedeli2012matter,abebe2013large}. These results agree with those obtained in \cite{munyeshyaka2025matter}. 
\end{itemize}
Motivated by the positive feedback of the considered models in the context of $f(Q,\mathcal{L}_{m})$ gravity, since the models present significant results at the perturbation level, we further analyse the implications of the considered models observationally, we use different observational datasets and MCMC analysis to constrain the model parameters. Using (i) CC+PP (ii) CC+DES (iii) CC+BAO+fs8 (iv) CC+BAO+fs8+PP (v) CC+BAO+fs8+DES, the resulting posterior distributions are visualized using triangular (corner) plots, generated by analysing the MCMC chains with \texttt{GetDist}. The marginalized parameter distribution is shown in Fig. (\ref{fig:lcdm_post}) and the corresponding best fit values are given in Tab. [\ref{tab:modelA_sn_params}], respectively for (i) CC+PP (ii) CC+DES. 
This table contains the mean values and corresponding errors. For this particular form of $f(Q,\mathcal{L}_{m})$ gravity model, we found that  for the pantheon data set including the CC data, $H_{0}=67.6$ km/s/Mpc--closely matching the plank collaboration results. Using two separate type $I_{a}$ Supernova, we recover $H_{0}$ value consistent with those from $\Lambda$CDM. The DES catalogue provides $H_{0}=69.05$ km/s/Mpc closer to the SHOES results. We notice that our findings are quite similar to the results reported in \cite{mhamdi2024constraints,sahlu2025structure} and \cite{mhamdi2024cosmological} for different gravity models considered.  After constraining the model parameters resulting from the modified Friedmann equation for the background, we extend our analysis to the perturbation level and we find the structure growth equation (eq. \ref{eq40}). This equation combined with $\sigma_8$ (eq. \ref{eq41}) enables us to constrain model parameters including $\sigma_8$. The obtained mean values and corresponding errors are presented in table (\ref{tab:modelA_fs8_params}) and the corner plot  for RSD data  is presented in fig. (\ref{fig:corner_mod1_sn}) . For the $\Lambda$CDM, we found that for the BAO and RSD data sets, $H_{0}=70.1$ km/s/MPC and $\sigma_8=0.771$, whereas for this particular form of $f(Q,\mathcal{L}_{m})$ gravity model $H_{0}=72.2$ km/s/MPC for CC+BAO+$f\sigma 8$, and nearly \(H_0 = 70.1\) km/s/Mpc for CC+BAO+$f\sigma 8$+PP and CC+BAO+$f\sigma 8$+DES combinations-higher than $\Lambda$CDM results.  The results can be put in comparison with the findings reported in \cite{panotopoulos2021growth,mhamdi2024constraints,mhamdi2024cosmological,sahlu2025structure} and in \cite{albuquerque2022designer} which we find no contradiction. An effective equation of state $\gamma=1.07$ deviates from pressure-less dark matter ($\gamma=1$) and this deviation do appear also  for CC+PP and CC+DES data sets, where $\gamma=1.03$. The obtained value of $\sigma_8$ shows noticeable deviation from the $\Lambda$CDM value. DES data set yields $\sigma8=0.658$ for model A, with a deviation from $\Lambda$CDM by around $2.14\sigma$--suggesting a potential resolution to the $\sigma_8$ tension in this $f(Q,\mathcal{L}_{m})$ gravity context. CC+BAO+$f\sigma 8$ combination yields $H_{0}=72.3$km/s/MPC, very close to the SHOES value. In this case $\gamma=1.127$ indicating a significant deviation from a pressure-less fluid.  The results show that the $f(Q,\mathcal{L}_{m})$ gravity fits the data well as compared with the $\Lambda$CDM parameter Values, indicating that this particular choice of $f(Q,\mathcal{L}_{m})=-\alpha Q+2l_{m}+\beta$  gravity model has potential to explain the dynamics of the universe including the cosmic accelerated phase and align well with observations at perturbation level unlike for the  $f=-\alpha Q+(2l_{m})^{2}+\beta$ model (Model B) which failed to fit the considered data sets. In conclusion, this type of phenomenology analysis of $f(Q,\mathcal{L}_{m})$ gravity models provides insight on the types of deviations that might be expected on cosmological observables and that can be used to distinguish the model from $\Lambda$CDM. Therefore it will be of interest to constrain different $f(Q,\mathcal{L}_{m})$ gravity models from combined data analysis of large scale structure, CMB, BAO and SN$I_{a}$ data sets. Work in this context is in progress.

\begin{acknowledgments}
	PKD wishes to acknowledge that part of the numerical computation of this work was carried out on the computing cluster Pegasus of IUCAA, Pune, India and PKD would like to acknowledge the Inter-University Centre for Astronomy and Astrophysics (IUCAA), Pune, India, for providing him a Visiting Associateship under which a part of this work was carried out. Also, PKD would like to thank the Isaac Newton Institute for Mathematical Sciences, Cambridge, for support and hospitality during the programme Statistical mechanics, integrability and dispersive hydrodynamics where work on this paper was undertaken. This work was supported by EPSRC grant no EP/K032208/1.
	AM acknowledges the hospitality of the University of Rwanda-College of Science and Technology, where part of this work was conceptualised and completed.
	SH acknowledges the support of National Natural Science Foundation of China under Grants No. W2433018 and No.
	11675143, and the National Key Reserach and development Program of China under Grant No. 2020YFC2201503.
	
\end{acknowledgments}

\nocite{*}


\end{document}